\documentclass[]{spie}  
\usepackage[]{graphicx}
\usepackage{amssymb}
\usepackage{amsbsy}
\usepackage{amsmath}
\usepackage{bm}

\title{ Quantum Electromagnetic Vacuum Fluctuations in Inhomogeneous Dielectric Media }

\author{ Shin-itiro Goto, Robin  W Tucker, Timothy J Walton 
\skiplinehalf
The Cockcroft
Institute and  Lancaster University, Lancaster, LA1 4YB, UK
}

\authorinfo{Further author information: (E-mail: r.tucker@lancaster.ac.uk, Telephone: +44 (0) 1524 593610}

\newcommand\PHI[1]{\boldsymbol\Phi_{#1}^T}
\newcommand\PSI[1]{\boldsymbol\Psi_{#1}^T}

\newcommand{\EM}{ electromagnetic }
\newcommand{\beq}{\begin{equation}}
\newcommand{\beqa}{\begin{eqnarray}}
\newcommand{\eeq}{\end{equation}}
\newcommand{\eeqa}{\end{eqnarray}}

\newcommand{\rr}{ \bm r}
\newcommand{\eps}{\epsilon}
\newcommand{\ee}{\bm e}
\newcommand{\EE }{\bm E}
\newcommand{\bb}{\bm b}
\newcommand{\hh}{\bm h}

\newcommand{\dd}{\bm d}

\newcommand{\BB }{\bm B}
\newcommand{\AAA}{\bm A}
\newcommand{\calA}{ {\cal A}}
\newcommand{\ww}{\wedge}
\newcommand{\kx}{  \frac{n_x\pi } {L_x }}
\newcommand{\ky}{  \frac{n_y\pi } {L_y }}
\newcommand{\NN}{{n_x,n_y,p}}
\newcommand{\calN}{{\cal N}}

\newcommand{\TE}{{ (+),\mbox{\tiny $T\!E$} }}
\newcommand{\TM}{{ (+),\mbox{\tiny $T\!M$} }}
\newcommand{ \aop} {{\bm a} }

\newcommand{\df}{\frac{}{}\!}
\newcommand{\sTE}{{ \mbox{\tiny $T\!E$} }}
\newcommand{\sTM}{{ \mbox{\tiny $T\!M$} }}
\def\[{\left[}
\def\]{\right]}

\begin{document}
  \maketitle

\begin{abstract}

 A new mathematical and computational technique for calculating quantum vacuum expectation values of energy and  momentum   densities associated with electromagnetic fields in bounded domains containing inhomogeneous media is discussed. This technique is illustrated by calculating the mode contributions to the difference in the vacuum force expectation between opposite ends of an inhomogeneous dielectric non-dispersive medium confined to a perfectly conducting rigid box.

\end{abstract}



\section{INTRODUCTION}
\label{sec:intro}  

Electromagnetic quantum fluctuation phenomena lie at the heart of many processes where the interface between classical and quantum physics plays a prominent role.  Among these one may cite areas of quantum optics, micro-cavity physics, micro-fluidics, photonic structures, early Universe cosmo-genesis, dark energy  and cold-atom technology.  In these systems one is often confronted with phenomena that interrelate classical continuum mechanics, classical electromagnetism, cavity quantum electrodynamics and fundamental issues relating to fluctuation-dissipation mechanisms \cite{Loudon,Milton} \!\!. In particular, dynamic (material) fluctuations induced by quantum fluctuations of the electromagnetic field have experimental consequences and offer an exciting opportunity to confront the limitations of basic theory with observable data.
In technology such fluctuations may manifest themselves as quantum induced stresses. For example Casimir stresses  cannot be ignored as nano-structures develop  ever smaller miniaturizations.
In micro-fluidics, physical processes can be confined to (deformable) micro-cavities that are guided by electromagnetic fields.  Such micro-laboratories offer new possibilities to explore cavity QED experimentally as well as enhancing the control features of micro-fluidic design. Indeed it has even  been suggested that chemical processes in such an environment may shed light on the mechanism that evolves inert matter into living cells.

The role of quantum fluctuations in determining the behavior of fabricated micro-structures is becoming increasingly important in a wide area of science and technology. Such fluctuations are also at the heart of many fundamental problems in physics ranging from the stability of fundamental constituents of matter to the lack of progress in unifying quantum field theory with gravitation. Many of the problems arise due to a lack of knowledge of basic interactions between fields and matter at some scale and the need to regularize current theories in order to make experimental predictions. For renormalizable theories such as QED in the vacuum these are remarkably accurate. However macroscopic predictions directly from QED  that are affected by the presence of bulk macroscopic matter and material interfaces can be made with far less confidence since they depend critically on both geometric and constitutive modeling at macroscopic scales \cite{Inui1,Inui2} \!\!. In particular the role of quantum states of the electromagnetic field on the detailed behavior of isolated closed material micro-domains of polarisable matter remains an unsolved problem \cite{Nester} \!\!.

The quantization of the \EM field in the presence of  media  satisfying linear electromagnetic constitutive relations relies on a knowledge of a regular basis of eigen-solutions to a modified Helmholtz type equation that determines the electromagnetic fields. If the  electromagnetic properties of the media are discontinuous across  surfaces in space such solutions must satisfy jump conditions across them  dictated by Maxwell's equations for the fields. The nature of the eigen-spectrum of the vacuum Helmholtz operator on time-parameterized differential forms on space is  determined by the global topology of spatial domains with boundaries.
Mathematical procedures exist for analyzing such problems using the Hodge-Weyl-Friedrichs decomposition of forms on manifolds with boundary. In principle they can be used  to construct a Hilbert space with a real basis of {\it transverse (divergence-less) } forms (i.e. in the kernel of the co-derivative $\delta$) satisfying Dirichlet and Neumann boundary conditions. The split of this space  into mutually orthogonal subspaces is responsible  for the classification of electromagnetic fields into TE, TM and TEM modes in certain domains.
 For example, if the vacuum of 3-dimensional Euclidean space is partitioned into interior and exterior regions  by a closed perfectly conducting boundary surface one may establish an ortho-normal  basis of transverse $1$-forms $\PHI N(\rr),\,\PSI N(\rr)$ satisfying
appropriate boundary conditions in each region. For a hollow closed cavity these basis modes can be defined in terms of the eigen-$1$-forms $\Phi_N(\rr)$ and $\Psi_N(\rr)$  of the Hodge-de Rham operator (Laplacian) $\Delta$  on forms in space satisfying different boundary conditions:
\begin{eqnarray*}
    {\Delta\Phi_N=\mu^2_N\Phi_N}, \quad  {\Delta\Psi_N=\lambda^2_N\Psi_N},
\end{eqnarray*}\vspace{-0.7cm}
\begin{eqnarray*}
    \PHI N \equiv \frac{1}{\mu^2_N} \delta\, d \Phi_N,\quad  \PSI N  \equiv\frac{1}{\lambda^2_N}\delta\,d \Psi_N, \quad \delta\PHI N=0,\quad \delta\PSI N=0
\end{eqnarray*}
where $\delta\equiv -\#\, d \, \#$ is the Euclidean exterior co-derivative on $1$-forms on a simply-connected domain ${\cal U}\in {\bf R}^3 $, $d$ denotes the exterior derivative, $-\Delta=d\delta + \delta\, d$  and $N$ labels a triplet of infinitely denumerable discrete numbers labelling the real non-zero eigen-values and associated eigen-forms. In a simply-connected domain such $1$-forms can be employed to represent the Coulomb gauge Maxwell vector-potential $1$-form ${\bm A}={\bm A}^{\sTE}+{\bm A}^{\sTM}$ where:
\begin{eqnarray*}
    {\bm A}^{\sTE}(t,\rr)=\sum_N {\cal A}^{\sTE}_N(t)\,  \PHI N(\rr), \qquad {\bm A}^{\sTM}(t,\rr)=\sum_N {\cal A}^{\sTM}_N(t)\, (\# d\PSI N(\rr))
\end{eqnarray*}
and $\#$ denotes the Euclidean Hodge map on forms in space.
The eigen-values $\mu_N,\lambda_N$ determine the normal-mode frequencies of electromagnetic fields in the cavity.
A similar analysis can be performed in the exterior region (which may be non-simply connected and involve TEM modes). The computation of the  eigenvalues   $\lambda_N, \,\mu_N$ is the key precursor to all quantum computations since they characterize the extrinsic domain geometry  and determine the spectral content of the infinite number of quantum oscillators that represent the electromagnetic field.
However it is often difficult to determine analytically  such bases for generic domains.
Furthermore if the partition involves electrically neutral domains containing  linear media that may be inhomogeneous, anisotropic, magneto-electric or dispersive this program involves a modified Helmholtz classical boundary value problem \cite{LeonBook,LeonInhom,LeonInhom2} \!\!.

In this paper the modification necessitated by the presence of an inhomogeneous but non-dispersive and non-conducting medium contained in a closed rectangular 3-dimensional perfectly conducting cavity is explored. This is a precursor to a regularization scheme needed to extract finite quantum expectation values for stresses in the medium induced by quantum electromagnetic fluctuations.

\section{Inhomogeneous Dielectric Media}
\label{ch1}

Consider a smooth open region of space containing a stationary non-dispersive medium characterized by an inhomogeneous permittivity scalar $\eps(\rr)$ and constant permeability scalar $\mu=\mu_0$.  Denoting time derivatives with an over-dot the classical source-free Maxwell system to be solved is:
\begin{alignat}{2}
    d\,\ee &=-\dot\BB, \qquad &\frac{1}{\mu} d\, \bb &=\eps \dot \EE \\
    d\,\BB &=0, \quad & d\,(\eps\EE)& =0 \\
    d\,\eps\ne 0, \,&\, d\,\mu = 0, \qquad &\dot\eps=0, \,&\,\dot\mu=0
\end{alignat}
where $\ee,\,\,\bb$ denote time dependent electric and magnetic 1-forms respectively and $\EE=\#\ee, \,\, \BB=\#\bb$.
If the (time-dependent) spatial 1-form $\AAA$ and spatial 0-form $\phi$ belong to a class of gauge-equivalent potentials defining the electric and magnetic fields by
\beqa \ee=-\dot \AAA - d\,\phi,  \qquad \bb=\# d\, \AAA \label{FIELDS}  \eeqa
then the above system reduces to
\begin{eqnarray*}
    \delta(\eps \dot\AAA) + \delta (\eps d\,\phi)  &=& 0 \\
    \delta d\,\AAA + \eps\mu(d\,\dot\phi + \ddot\AAA) &=& 0
\end{eqnarray*}
In a particular gauge with $$\delta\,(\eps\AAA)=0$$ 
the equation for the scalar potential decouples:
\begin{eqnarray*}
    \delta d\,\AAA+ \eps\mu \ddot\AAA &=& -\eps\mu d\,\dot\phi \\
    \delta (\eps d\,\phi) &=& 0 .
\end{eqnarray*}
Furthermore, we may set $d\,\phi=0$ for systems without free charge\cite{Glauber} \!\!. Hence in this gauge the local Maxwell system above is solved in terms of spatial 1-forms satisfying $\delta (\eps\AAA)=0$ and the  equation:
\beqa  \delta d\,\AAA+\eps\mu\ddot\AAA=0. \label{HELM} \eeqa
This gives rise to a modified Helmholtz equation for time harmonic fields.
Across any non-conducting interface where the dielectric scalar is discontinuous one has two conditions on $\ee$ and $\bb$ restricted to the interface. At each point on the interface  the jump in the normal component of $\bb$ and the tangential component of $\ee$ must vanish. Furthermore if there are no real charges or electric currents on the interface the jump in  the normal component of  $\dd$ where  $\dd=\eps \ee$ and the tangential component of $\hh$  where $\hh=\mu^{-1}\bb$ must also vanish on the interface. If the interface is perfectly conducting one assumes that all fields vanish on the side of the interface that is free of the material medium when calculating the jump. It is worth noting that in a bounded inhomogeneous medium where $\eps$ is a continuous function of position one cannot exploit translational invariance in space and spatial Fourier transforms\cite{Brevik} to simply express normal modes in terms of eigen-forms of $ {\bm R}^3 $ spatial translation operators.

If a bounded domain $U \subset {\bm R}^3 $ contains a  dielectric  with a piecewise inhomogeneous permittivity one writes the general real 1-form solution  to (\ref{HELM})  on $U$ as
\beqa  \AAA(t,\rr)=\sum_N  \calA_N(t,\rr) \label{GEN}\eeqa
where $N$ denotes a triple of discrete labels. Suppose the dielectric is composed of $M$ sub-domains where the permittivity is $\eps_m(\rr)$ in the $m$-th sub-domain. Thus
\begin{eqnarray*}
\eps(\rr)=\sum_{m=1}^M \eps_m(\rr) {\cal Y}_m(\rr)
\end{eqnarray*}
where ${\cal Y}_m(\rr)$ is unity in the subdomain $U_{m}\subset U$ and zero elsewhere. For stationary electrically neutral dielectrics in domains $U$  bounded by conducting surfaces the electromagnetic jump conditions at interfaces above yield a homogeneous system of equations that determine a collection of eigen-modes (up to normalization) and the associated eigen-frequencies $\omega_N$. The number of distinct eigen-spaces and the degeneracies of the associated eigen-frequencies  depends on the rank  and symmetry of the homogeneous system which in turn reflects how the boundary geometry and boundary conditions affect the nature of the global topology of the domain. For a rank $S$ system  the eigen-modes may be written:
\beqa {\cal A}_N(t,\rr)=\sum_{s=1}^S\sum_{m=1}^M \left(
{\cal A}^{(+),s,m}_N(\rr){\cal Y}_m(\rr)\, e^{-i \omega_N^s t} + {\cal A}^{(-),s,m}_N(\rr){\cal Y}_m(\rr) \,e^{i \omega_N^s t}\right)
 \eeqa
where the $ \{{\cal A}^{(+),s,m}_N(\rr)\}=   \{{\cal A}^{(-),s,m }_N(\rr)\}^{*}      $ constitute a basis of solutions to (\ref{HELM}) subject to  $ \delta (\eps{\cal A}^{(+),s,m  }_N ) =0$ and the above jump conditions in the domain $U$.

\section{Electromagnetic Energy and Stresses}

\label{ch2}

Classical forces (and torques) transmitted by electromagnetic fields through the vacuum can be encoded into a covariant stress-energy-momentum tensor. The modification of such  a tensor for fields in a medium has long been a subject of debate and experimental investigation. This debate has continued when the fields become operators  subject to quantum laws. In this article  we adopt  a symmetrization of  the stress-energy-momentum tensor for media at rest advocated by Minkowski.  In terms of electromagnetic fields in any sub-domain  $U_m$ the instantaneous  classical electromagnetic energy is:
\beqa {\cal E}_m=\int_{U_m} \frac{1}{2}\left(  \ee \wedge \# \dd + \bb \wedge \# \hh \right) \label{ENERGY}.\eeqa

The component of the instantaneous integrated electromagnetic stress \cite{TuckerWalton} \!\!,  in a direction defined by a unit spacelike Killing vector field $K$ that generates spatial translations, transmitted across the side of any portion $\Sigma_m$  of an oriented surface in $U_m$ adjacent to the fields in the following integrand, is the force component:
\begin{eqnarray*}\label{FORCE}
    {\cal F}_{K,m}=\frac{1}{2}\int_{\Sigma_m}\left(  i_K\#\hh \ww  \bb  - \ee \ww i_K \# \dd - \# \bb \ww \hh(K) - \dd(K) \ww \# \ee\right)
\end{eqnarray*}

\subsection{Computation of Induced Dielectric Stresses by Electromagnetic Mode Fluctuations in a Cavity}

\label{ch3}

The above generalities will now be illustrated for a system comprised  of a simply-connected inhomogeneous dielectric medium bounded by a perfectly conducting stationary, inextensible rectangular box with sides of length $L_x,L_y,L_z$.  Finding non-trivial exact analytic solutions to (\ref{HELM}) is non-trivial. However, if $\eps(x,y,z)=\epsilon_0 \,\beta \exp(\alpha z/L_z)$ in Cartesian coordinates with real dimensionless positive constants $\beta,\alpha$, then general solutions satisfying the above boundary conditions can be expressed in terms of Bessel  and trigonometric functions. Since the interior $U$ of the box is simply connected the boundary conditions yield $S=2$ and a decomposition  into orthogonal  $TE$ and $TM$ modes with respect to the $z$-axis  is possible.  With opposite faces of the box  at $z=0$ and $z=L_z$ respectively the $TE$ mode structure is given in the above gauge by:
\begin{eqnarray*}
    {\cal A}_N^\TE(\rr)=\frac{ {\calN}_N^{\sTE}  }{\epsilon_0}\Phi_N^{\sTE}\!\[\eta^{\sTE}_N(z)\] \left\{ \df k_x \sin(k_x x)\cos(k_y y) d\,y -k_y \cos(k_x x) \sin(k_y y) d\, x \right\}
\end{eqnarray*}
where $k_x=\kx, k_y=\ky$, $N$ stands for the  triple $(\NN)$ with  $n_x,n_y$  positive integers (including zero) and ${\calN}^{\sTE}_N$
denotes  a normalization constant.  Furthermore
\begin{eqnarray*}
    \Phi_N^{\sTE}\[\eta^{\sTE}_N(z)\] &=& J_{\nu_N^{\sTE} }\!\[ \eta^{\sTE}_N(z)\] +  \zeta_N^{\sTE}   Y_{\nu_N^{\sTE} }\!\[ \eta^{\sTE}_N(z)) \]
\end{eqnarray*}
where
\begin{eqnarray*}
    \eta_N^{\sTE}(z)= \frac{ 2 L_z \omega_N^{\sTE} \sqrt{\beta} \exp( \frac{ \alpha z} {2 L_z } ) }  { \alpha c_0 }, \quad \nu_N^{\sTE}= \frac{2 L_z } { \alpha}\sqrt{ k_x^2 +k_y^2 }, \quad \zeta_N^{\sTE}= -\frac{J_{\nu_N^{\sTE} }\!\[ \eta^{\sTE}_N(0)\]  }  { Y_{\nu_N^{\sTE} }\!\[ \eta^{\sTE}_N(0)\]  }
\end{eqnarray*}
with $c_{0}^{2}=\frac{1}{\epsilon_{0}\mu_{0}}$ in these expressions and the $\omega_N^{\sTE}$ are the values of the $p$-th roots of the $TE$-mode spectrum generating equation:
\begin{eqnarray*}
    J_{ \nu_N^{\sTE} }\!\[\eta^{\sTE}_N(0)\] Y_{ \nu_N^{\sTE}} \!\[ \eta^{\sTE}_N(L_z)\]  - J_{ \nu_N^{\sTE} }\!\[ \eta^{\sTE}_N(L_z)\] Y_{ \nu_N^{\sTE} }\!\[ \eta^{\sTE}_N(0)\] &=& 0.
\end{eqnarray*}
The $TM$ mode structure is given by:
\begin{eqnarray*}
{\cal A}_N^\TM(\rr) &=& \frac{ \calN_N^{\sTM} \omega_N^{\sTM} }{\epsilon_0 c_0}  \left\{   \left( \Phi_N^{\prime \sTM}\!\[\eta^{\sTM}_N(z)\]
  +  \frac{ \Phi_N^{ \sTM}\!\[\eta^{\sTM}_N(z)\] } {\eta^{\sTM}_N(z) }  \right)
      \left( \df k_x \cos(k_x x)\sin(k_y y) d\,x +k_y \sin(k_x x) \cos(k_y y) d\, y\right) \right. \\
          && \hspace{2.5cm} \left. +  \frac{ \alpha}{2L_z \eta_N^{\sTM}} \left( \df (\nu_N^{\sTM})^2-1  \right) \Phi_N^{\sTM}\!\[ \eta^{\sTM}_N(z)\]  \sin(k_x x) \sin(k_y y)\,d\,z )      \right\}
\end{eqnarray*}
with normalization constant $\calN^{\sTM}_N$,
\begin{eqnarray*}
    \Phi_N^{\sTM}\[ \eta^{\sTM}_N(z)  \] &=&  J_{ \nu_N^{\sTM} }\!\[\eta^{\sTM}_N(z)\]  +  \zeta_N^{\sTM}   Y_{ \nu_N^{\sTM} }\!\[ \eta^{\sTM}_N(z) \]
\end{eqnarray*}
where
\begin{eqnarray*}
    \eta_N^{\sTM}(z)= \frac{ 2L_z \omega_N^{\sTM} \sqrt{\beta} \exp( \frac{ \alpha z}{2 L_z } ) }{ \alpha c_0 }, \quad (\nu_N^{\sTM})^2= \frac{4 L_z^2 } { \alpha^2}{( k_x^2 +k_y^2) }   + 1, \quad
    \zeta_N^{\sTM}=- \frac{ \eta^{\sTM}_N(0)  J_{ \nu_N^{\sTM} }^{\prime}\!\[ \eta^{\sTM}_N(0)\]     + J_{ \nu_N^{\sTM} }\!\[ \eta^{\sTM}_N(0) \]}
    {\eta^{\sTM}_N(0)  Y_{ \nu_N^{\sTM} }^{\prime}\!\[ \eta^{\sTM}_N(0)\]     + Y_{ \nu_N^{\sTM} }\!\[ \eta^{\sTM}_N(0) \]   }.
\end{eqnarray*}
In this case the $\omega_N^{\sTM}$ are the values of the $p$-th roots of the $TM$-mode spectrum generating equation which may be written in the form:
\begin{eqnarray*}
  \widetilde J_{ \nu_N^{\sTM} }\!\[ \eta^{\sTM}_N(0)\]\widetilde  Y_{ \nu_N^{\sTM} }\!\[ \eta^{\sTM}_N(L_z)\] - \widetilde J_{ \nu_N^{\sTM} }\!\[ \eta^{\sTM}_N(L_z)\] \widetilde Y_{ \nu_N^{\sTM} }\!\[ \eta^{\sTM}_N(0)\]  &=& 0
\end{eqnarray*}
where $\widetilde f(\eta)\equiv \eta f^\prime(\eta) + f(\eta) $ for any $f(\eta)$.
These expressions enable one to calculate  the field modes for  $\ee,\bb,\hh,\dd$  using  (\ref{FIELDS}) and the constitutive relations.

The quantum description can be constructed by generalizing the methods used in vacuum cavity QED. A Fock space of quantised modes is introduced by introducing the  annihilation  and creation operators $  \aop_N^s $   and $ {\aop_{N^\prime}^{s^\prime}}^{\dagger}  $   satisfying the commutator relations:
$$[  \aop_N^s, {\aop_{N^\prime}^{s^\prime}}^{\dagger} ]=\delta_{ N { N^\prime} }\, \delta ^{ s {s^\prime} }$$
for $s\in\{TE,TM\}$.
The Fock space vacuum state $\Lambda_0  $ is annihilated by all $  \aop_N^s $.
In the above gauge stationary quantum modes in a closed cavity are described by the Hermitian operator
\begin{eqnarray*}
    \widehat{\cal A}_N(t,\rr)=\sum_{s\in\{TE,TM\}}\sum_{m=1}^M \left( {\aop_{N^\prime}^{s^\prime}}^{\dagger}
 {\cal A}^{(+),s,m}_N(\rr){\cal Y}_m(\rr)\, e^{-i \omega_N^s t} +  \aop_N^s {\cal A}^{(-),s,m}_N(\rr){\cal Y}_m(\rr) \,e^{i \omega_N^s t}\right).
\end{eqnarray*}
The quantum field modes for the Hermitian operators $\widehat\ee,\widehat\bb,\widehat\hh,\widehat\dd$  follow from   (\ref{FIELDS})  but with this mode operator and the corresponding operator constitutive relations. Replacing the classical fields  for the dielectric filled cavity in the classical expression for   ${\cal E}$ by such operators yields the quantum hamiltonian $\widehat{\cal E}$ for quantum fields in the cavity. Its expectation value $ E_{\Lambda_0}[  \widehat{\cal E} ] $ in the Fock space vacuum state requires renormalization \cite{Santos,BordagBook} \!\!. This is effected by subtracting from an infinite mode sum an expectation value of the energy of a system with a homogeneous medium. To effect this subtraction both mode summations generally require a regularization scheme for their definition.  Thus for the system with conducting boundaries  and a discrete spectrum one defines:
\begin{eqnarray*}
    \langle {\cal E}\rangle_{\text{reg}} &\equiv&  \frac{\hbar }{2} \sum_{s \in \{TE,TM\}}\sum_N \omega_N^s \, \psi(\kappa,\omega_N^s)
\end{eqnarray*}
for some suitable smooth function satisfying  $\psi(0,\omega_N^s)=1$ that renders the summations meaningful.
Each  cavity mode labeled by $N,s$,  with eigen-frequency $\omega_N^s$, contributes the factor $ \frac{\hbar}{2}\, \omega_N^s$ to the vacuum expectation of the (regularized) energy.
The well-defined condition
\begin{eqnarray*}\label{QENERGY}
    {\cal E}_N^s\equiv \int_{U} \frac{1}{2}\left( \df \ee_N^s \wedge \# \dd_N^s + \bb_N^s \wedge \# \hh_N^s \right) \;\;=\;\; \frac{\hbar}{2} \;\omega_N^s
\end{eqnarray*}
fixes the normalizations  ${\cal N }_N^s$ of the mode amplitudes ${\calA}_N^{(+), s}$ and their conjugates:
\begin{eqnarray*}
    ({\cal N}_N^{\sTE})^2 &=& \frac{16 \hbar \epsilon_0 }{\alpha^{2}\sqrt{\beta} c_{0}({\nu_N^{\sTE}})^2}\frac{L_z^2}{L_x L_y}\frac{1}{I^{\sTE}_{N}\Omega^{\sTE}_{N}}
\end{eqnarray*}
 where, with $\Omega^{\sTE}_N=\frac{2 \omega_N^{\sTE} L_z \sqrt{\beta}  } { \alpha\, c_0} $,
\begin{eqnarray*}
    I^{\sTE}_{N} &=& e^\alpha \left( {\Phi^{\prime}}^{\sTE}_N\!\[\Omega^{\sTE}_N e^{\frac{\alpha}{2}}\] \right)^2 - \left( {\Phi^{\prime}}^{\sTE}_N\!\[ \Omega^{\sTE}_N\] \right)^2
\end{eqnarray*}
and
\begin{eqnarray*}
    ({\cal N}_N^{\sTM})^2 &=& \frac{64\hbar \epsilon_{0}\sqrt{\beta}}{\alpha^{4}c_{0}[({\nu_N^{\sTM}})^2-1]} \frac{ L_z^4}{L_x L_y}  \frac{1}{I^{\sTM}_{N}\Omega_N^{\sTM}}
\end{eqnarray*}
where, with $\Omega^{\sTM}_{N}=\frac{2 \omega_N^{\sTM} L_z \sqrt{\beta}  } { \alpha\, c_0}  $,
\begin{eqnarray*}
    I^{\sTM}_{N} &=&  \left(\df 1 - (\nu^{\sTM}_{N})^{2} + (\Omega^{\sTM}_{N})^{2}e^{\alpha} \right) \left( \df \Phi^{\sTM}_{N}\!\[\Omega^{\sTM}_{N}e^{\frac{\alpha}{2}}\]\right)^{2} - \left(\df 1 - (\nu^{\sTM}_{N})^{2} + (\Omega^{\sTM}_{N})^{2} \right)\left(\df \Phi^{\sTM}_{N}\!\[\Omega^{\sTM}_{N}\]\right)^{2}.
\end{eqnarray*}
The mode contributions to the vacuum expectation values of the induced electromagnetic stress field  in the dielectric can now be calculated from the Hermitian  operator-valued stress 2-form:
\beqa \widehat{\cal \sigma}_{K}=\frac{1}{2}\left(  i_K\#\widehat\hh \ww  \widehat\bb  - \widehat\ee \ww i_K \# \widehat\dd - \# \widehat\bb \ww \widehat\hh(K) -\widehat \dd(K) \ww \# \widehat\ee\right). \label{QFORCE}  \eeqa
The quantum expectation value of the regularized force component in the Fock vacuum state $\Lambda_0$ acting perpendicular to a surface $\Sigma_{z_{0}}$ of the box at $z=z_{0}$ is
\begin{eqnarray*}
    \langle  {\cal F}_{z_{0}} \rangle_{\text{reg}} &\equiv& \left. E_{\Lambda_0}\left[ \int_{\Sigma_{z_{0}}}\, \widehat \sigma_{ \frac{ \partial} { \partial z}} \right]_{\text{reg}} \right|_{z=z_{0}}.
\end{eqnarray*}
In a box containing a homogeneous permittivity, the expectation values of the force at the end faces of the box at $z=0,z=L_z$ are equal. This is not the case for an inhomogeneous dielectric in general. Indeed one finds, after some calculation, the difference between the force expectations
\begin{eqnarray*}
    \langle \Delta {\cal F} \rangle_{\text{reg}}  \;\;\equiv\;\; \langle  {\cal F}_{0} \rangle_{\text{reg}}  -  \langle {\cal F}_{L_z} \rangle_{\text{reg}}   &=& \frac{ \hbar \alpha} { 4 L_z}  \sum_{s \in \{ TE,TM\}} \sum_N \omega_N^{s} \, \psi(\kappa,\omega_N^s)
\end{eqnarray*}
This is a surprisingly simple result given the complexity of the mode structures involved. To effect a renormalization of this result requires a computation which will not be reported here.

\section{Conclusion}

\label{ch4}

It is expected that a non-zero $ \langle \Delta {\cal F} \rangle $ will survive renormalization when the regulator is removed and this indicates that any  confined inhomogeneous material dielectric  must sustain stresses induced by electromagnetic quantum fluctuations if the confining domain is rigid. If the medium remains static such stresses  induce mechanical (elastic) stresses in the dielectric to maintain equilibrium. Unlike similarly induced classical stresses by the classical gravitational field in the laboratory (that vary with the orientation of the dielectric) the quantum induced electromagnetic stresses are permanent. In principle they could be detected experimentally by noting the variation of the induced stress field within the dielectric with variations of the permittivity  inhomogeneities. Such variations might be detected  using photo-elastic effects on the polarization of light passing through the medium.

\acknowledgements
The authors are grateful to STFC and the EPSRC  for financial support for this research which is part of the Alpha-X collaboration.

\nocite{*}
\bibliography{report}

\begin{thebibliography}{10}

\bibitem{Loudon}
Loudon, R., ``{T}heory of the {R}adiation {P}ressure on {D}ielectric
  {S}urfaces,'' {\em J. Mod. Opt.}~{\bf 49},  821--838 (2002).

\bibitem{Milton}
Milton, K.~A., ``{R}esource {L}etter: {V}an der {W}aals and {C}asimir-{P}older
  {F}orces,'' {\em arXiv:1101.2238v1} ,  1--37 (2011).

\bibitem{Inui1}
Inui, N., ``{C}asimir {E}nergy of the {E}vanescent {F}ield between
  {I}nhomogeneous {D}ielectric {S}labs,'' {\em J. Phys. Soc. Jpn.}~{\bf 77},
  084001:1--8 (2008).

\bibitem{Inui2}
Inui, N., ``{A} {G}eneralized {M}ode {S}ummation {F}ormula of the
  {Z}ero-{P}oint {E}nergy in a {C}avity,'' {\em J. Phys. Soc. Jpn.}~{\bf 72},
  1035--1040 (2003).

\bibitem{Nester}
Nesterenko, V.~V., Lambiase, G., and Scarpetta, G., ``{C}alculation of the
  {C}asimir {E}nergy at {Z}ero and {F}inite {T}emperature: {S}ome {R}ecent
  {R}esults,'' {\em Riv. Nuovo Cimento}~{\bf 27},  1--74 (2004).

\bibitem{LeonBook}
Leonhardt, U.,  [{\em {E}ssential {Q}uantum {O}ptics: from {Q}uantum
  {M}easurements to {B}lack {H}oles}{\nolinebreak\hspace{0.1em}]}, Cambridge
  University Press, Cambridge (2010).

\bibitem{LeonInhom}
Xiong, C., Kelsey, T., Linton, S., and Leonhardt, U., ``{T}owards the
  {C}alculation of {C}asimir {F}orces for {I}nhomogeneous {P}lanar {M}edia,''
  {\em submitted} ,  1--9.

\bibitem{LeonInhom2}
Philbin, T.~G., Xiong, C., and Leonhardt, U., ``{C}asimir {S}tress in an
  {I}nhomogeneous {M}edium,'' {\em Ann. Phys.}~{\bf 325},  579--595 (2009).

\bibitem{Glauber}
Glauber, R.~J. and Lewenstein, M., ``{Q}uantum {O}ptics of {D}ielectric
  {M}edia,'' {\em Phys. Rev. A: At. Mol. Opt. Phys.}~{\bf 43},  467--491
  (1991).

\bibitem{Brevik}
Brevik, I. and Sollie, R., ``{C}asimir {F}orce on a {S}pherical {S}hell when
  $\epsilon(\omega)\mu(\omega)=1$,'' {\em J. Math. Phys.}~{\bf 31},  1445--1455
  (1990).

\bibitem{TuckerWalton}
Tucker, R.~W. and Walton, T.~J., ``{A}n {I}ntrinsic {A}pproach to {F}orces in
  {M}agnetoelectric {M}edia,'' {\em Il Nuovo Cimento C}~{\bf 32},  205--229
  (2009).

\bibitem{Santos}
Santos, F.~C., Sobrinho, J. J.~P., and Tort, A.~C., ``{E}lectromagnetic {F}ield
  {C}orrelators, {M}axwell {S}tress {T}ensor, and the {C}asimir {E}ffect for
  {P}arallel {W}alls,'' {\em Braz. J. Phys.}~{\bf 35},  657--666 (2005).

\bibitem{BordagBook}
Bordag, M., Klimchitskaya, G.~L., Mohideen, U., and Mostepanenko, V.~M.,  [{\em
  {A}dvances in the {C}asimir {E}ffect}{\nolinebreak\hspace{0.1em}]}, Oxford
  University Press, Oxford (2009).

\end{thebibliography}
\bibliographystyle{spiebib}

\end{document}